\documentclass[aps,prb,twocolumn,superscriptaddress]{revtex4-2}

\usepackage{graphicx, color, xcolor}
\usepackage{amsmath,amsfonts,amssymb}
\usepackage{booktabs} %
\usepackage{soul}
\usepackage{hyperref}
\hypersetup{
    colorlinks=true,
    linkcolor=blue,
    filecolor=magenta,      
    urlcolor=cyan,
    pdftitle={Evaluating radiation impact on transmon qubits in above and underground facilities},
    }
    
\usepackage[heightadjust=all]{floatrow}
\usepackage{comment}

\newcommand{\beginsupplement}{\setcounter{table}{0}  \renewcommand{\thetable}{S\arabic{table}} \renewcommand{\theHtable}{S\arabic{table}} \setcounter{figure}{0} \renewcommand{\thefigure}{S\arabic{figure}} \renewcommand{\theHfigure}{S\arabic{figure}}}

\providecommand{\ignore}[1]{}

\newcommand{\ket}[1]{|#1\rangle}

\def\lsim{\mathrel{\rlap{\lower4pt\hbox{\hskip1pt$\sim$}}
    \raise1pt\hbox{$<$}}}                
\def\gsim{\mathrel{\rlap{\lower4pt\hbox{\hskip1pt$\sim$}}
    \raise1pt\hbox{$>$}}}                

\begin{document}

\title{Evaluating radiation impact on transmon qubits in above and underground facilities}

\author{Francesco De Dominicis}
\email{Equal contribution}
\affiliation{Gran Sasso Science Institute, L’Aquila I-67100, Italy}
\affiliation{INFN – Laboratori Nazionali del Gran Sasso, Assergi (L’Aquila) I-67100, Italy}

\author{Tanay Roy$\ ^*$}
\email{Corresponding author: roytanay@fnal.gov}
\affiliation{Superconducting Quantum Materials and Systems Center (SQMS), Fermi National Accelerator Laboratory, Batavia, IL 60510, USA}

\author{Ambra Mariani}
\email{Corresponding author: ambra.mariani@roma1.infn.it}
\affiliation{INFN -- Sezione di Roma, Roma I-00185, Italy}

\author{Mustafa Bal}
\affiliation{Superconducting Quantum Materials and Systems Division, Fermi National Accelerator Laboratory (FNAL), Batavia, IL 60510, USA}

\author{Camilla Bonomo}
\affiliation{INFN -- Sezione di Roma, Roma I-00185, Italy}
\affiliation{Dipartimento di Fisica, Sapienza Universit\`a di Roma, Roma I-00185, Italy}

\author{Nicola Casali}
\affiliation{INFN -- Sezione di Roma, Roma I-00185, Italy}

\author{Ivan Colantoni}
\affiliation{Consiglio Nazionale delle Ricerche, Istituto di Nanotecnologia}
\affiliation{INFN -- Sezione di Roma, Roma I-00185, Italy}

\author{Francesco Crisa}
\affiliation{Illinois Institute of Technology}

\author{Angelo Cruciani}
\affiliation{INFN -- Sezione di Roma, Roma I-00185, Italy}

\author{Fernando Ferroni}
\affiliation{Gran Sasso Science Institute, L’Aquila I-67100, Italy}
\affiliation{INFN -- Sezione di Roma, Roma I-00185, Italy}

\author{Dounia L Helis}
\affiliation{INFN – Laboratori Nazionali del Gran Sasso, Assergi (L’Aquila) I-67100, Italy}

\author{Lorenzo Pagnanini}
\affiliation{Gran Sasso Science Institute, L’Aquila I-67100, Italy}
\affiliation{INFN – Laboratori Nazionali del Gran Sasso, Assergi (L’Aquila) I-67100, Italy}

\author{Valerio Pettinacci}
\affiliation{INFN -- Sezione di Roma, Roma I-00185, Italy}

\author{Roman Pilipenko}
\affiliation{Superconducting Quantum Materials and Systems Division, Fermi National Accelerator Laboratory (FNAL), Batavia, IL 60510, USA}

\author{Stefano Pirro}
\affiliation{INFN – Laboratori Nazionali del Gran Sasso, Assergi (L’Aquila) I-67100, Italy}

\author{Andrei Puiu}
\affiliation{INFN – Laboratori Nazionali del Gran Sasso, Assergi (L’Aquila) I-67100, Italy}

\author{Alberto Ressa}
\affiliation{INFN -- Sezione di Roma, Roma I-00185, Italy}

\author{Alexander Romanenko}
\affiliation{Superconducting Quantum Materials and Systems Division, Fermi National Accelerator Laboratory (FNAL), Batavia, IL 60510, USA}

\author{Marco Vignati}
\affiliation{INFN -- Sezione di Roma, Roma I-00185, Italy}
\affiliation{Dipartimento di Fisica, Sapienza Universit\`a di Roma, Roma I-00185, Italy}

\author{David v Zanten}
\affiliation{Superconducting Quantum Materials and Systems Division, Fermi National Accelerator Laboratory (FNAL), Batavia, IL 60510, USA}

\author{Shaojiang Zhu}
\affiliation{Superconducting Quantum Materials and Systems Division, Fermi National Accelerator Laboratory (FNAL), Batavia, IL 60510, USA}

\author{Anna Grassellino}
\affiliation{Superconducting Quantum Materials and Systems Division, Fermi National Accelerator Laboratory (FNAL), Batavia, IL 60510, USA}

\author{Laura Cardani}
\affiliation{INFN -- Sezione di Roma, Roma I-00185, Italy}

\date{\today}

\begin{abstract}
Superconducting qubits can be sensitive to abrupt energy deposits caused by cosmic rays and ambient radioactivity. 
While previous studies have explored correlated effects in time and space due to cosmic ray interactions, we present the first direct comparison of a transmon qubit’s performance measured at two distinct sites: the above-ground SQMS facility (Fermilab, US) and the deep-underground Gran Sasso Laboratory (Italy). Despite the stark difference in radiation levels, we observe a similar average qubit relaxation time of approximately 80 microseconds at both locations.
To further investigate potential radiation-induced events, we employ a fast decay detection protocol, comparing the relative rates of triggered events between the two environments. Although intrinsic noise remains the dominant source of single errors in superconducting qubits, our analysis revealed a significant excess of radiation-induced events for high-coherence transmon qubits operated above-ground.  Finally, using $\gamma$-ray sources with increasing activity levels, we evaluate the qubit response in a controlled low-background environment.
\end{abstract}

\maketitle
Over the past decades, superconducting circuits have emerged as a leading technology for applications in 
quantum information~\cite{Barends2014, Arute:2019, Kjaergaard2020, RigettiAspen, wu:2021, IBM:eagle}, photon, and  particle detection~\cite{Dixit2021DM, Agrawal2023DM, Day2003, Irwin:1995, Ullom_2015, Berggren:2023, Cruciani_2018, bullkid_2022, Mazin:12, Monfardini_2011}. In  recent years, several research groups have started investigating the impact of particle interactions on the performance of superconducting quantum bits (qubits) with different geometries and materials~\cite{demetra:2020, mcewen2024gap_engineering}. For example, it has been observed that the lifetime of a transmon qubit can decrease when exposed to high levels of radiation~\cite{Vepsalainen:2020}. Additionally, the effect of high energy deposits can rapidly propagate from the initial location of deposition, inducing correlated errors or even general system failure~\cite{McEwen:2021, Li2024direct}. Wilen et al.~\cite{Wilen:2021} found that the rate of charge jumps in an array of charge-sensitive qubits (operated as electrometers) was consistent with the expected rate of interactions from ambient radioactivity and cosmic rays. In contrast, Thorbeck et al.~\cite{Thorbeck:2022} 
observed no significant direct effects from radiation,
but suggested that ionizing radiation may interact with two-level systems (TLSs), a dominant noise source in superconducting qubits~\cite{Klimov:2018, deGraaf:2017, Burnett:2014, Bilmes:2020, Lisenfeld:2019, Martinis:2005, Muller_2019, McRae_2020}.

Other studies have shown that suppressing radioactivity can enhance the performance of superconducting circuits. 
Operating devices in low-background facilities reduces correlated errors~\cite{Bratrud2024FNAL}, stabilizes fluxonium qubits~\cite{Gusenkova:2022}, and improves the internal quality factor of superconducting resonators~\cite{Cardani:2021}. The latter publication also determined that the substrate on which qubits are deposited, rather than the qubits themselves, is the primary target for particle interactions. 

These findings have motivated efforts in further directions aimed at studying and mitigating possible effects of radioactivity and cosmic rays. Promising results have been achieved by equipping the chip with ``traps" designed to absorb phonons produced (also) by radioactive interactions~\cite{Henriques:2019, martinis:2021, iaia_2022}, and using gap engineering~\cite{kamenov2023gap_engineering, mcewen2024gap_engineering, Harrington:2024}. 
Other approaches focus on minimizing radiation from qubit construction materials, such as Printed Circuit Boards (PCBs), which have been shown to produce a sizeable rate of events in the qubit chip~\cite{cardani:2023}. Furthermore, it is also worth mentioning recent proposals of equipping qubits with particle detection systems~\cite{orrell2021sensor} or external muon detectors~\cite{Harrington:2024,Li2024direct,mariani2023}.

In this work, we present the first study of a charge-insensitive transmon qubit with energy relaxation time $\text{T}_1$ of the order of 0.1\,milliseconds, measured in two laboratories with markedly different radiation levels: the \emph{Quantum Garage} of the Superconducting Quantum Materials and Systems (SQMS) Center at Fermilab (FNAL, Illinois, US), and the \emph{Ieti Facility} located in the deep-underground Gran Sasso Laboratory of INFN (LNGS, L'Aquila, Italy). The aim of this study is to quantify the effect of environmental radioactivity on such a device.

The paper is structured as follows. To understand how different radiation environments affect qubit performance, first we compare the average relaxation time of a qubit above ground at FNAL and deep underground at LNGS (Sections~\ref{sec:device and locations} --~\ref{sec:comparison}). Our results show that, as expected, radioactivity has a negligible effect even on transmons with a relaxation time of 0.1 milliseconds. By developing a fast decay detection with active reset protocol and a novel analysis technique, we are able to disentangle radiation-induced events (Sections~\ref{sec:detection protocol} --~\ref{sec:selection}).  We systematically explore how various experimental parameters---such as qubit relaxation time and the sampling period of the fast reset---influence the results. We find that radiation induced event rates are significantly higher at FNAL compared to that at LNGS. Finally, we expose the superconducting chip to calibrated $\gamma$-ray sources in a controlled, low-background environment, to further characterize the qubit's response to radiation and validate our findings (Section~\ref{sec:results}).

\begin{figure}[t]
\includegraphics[width=\columnwidth]{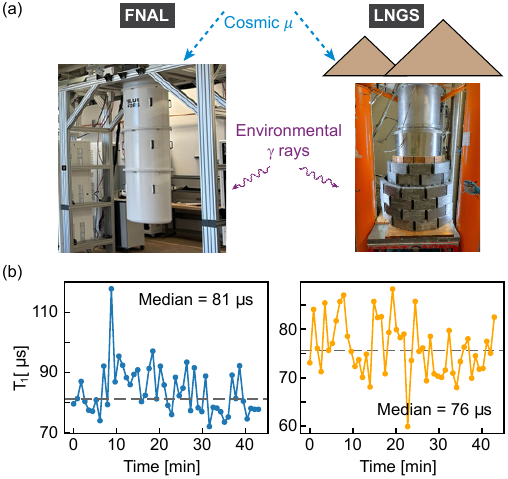}
\caption{\textbf{Comparison of the two experimental sites.} (a) The above-ground cryostat at FNAL is exposed to environmental radiation and cosmic-rays, whereas the deep-underground cryostat at LNGS is strongly guarded from cosmic muons by the 1.4 km deep rock overburden and from environmental $\gamma$ ray through a Copper plus Lead shielding. (b) The left and right panels display data taken at FNAL and LNGS, respectively. Standard $\text{T}_1$ measurements performed on the same device show similar fluctuations at both sites. However, these measurements are too slow to detect rapid T$_1$ drops at millisecond time scales potentially due to radiation impact.}
\label{fig:compare} 
\end{figure}

\section{Device and Experimental locations}
\label{sec:device and locations}
The chip consisted of a 432\,$\mu$m thick HEMEX grade Sapphire substrate with dimensions of 7.5$\times$7.5\,mm$^2$. The entire substrate was covered in Niobium, which creates a ground plane that may act as a phonon trap, diminishing phonon propagation and absorption in the active part of the qubit~\cite{yelton2024modeling, martinez2019}. Eight Niobium transmons with different geometries were deposited on the chip, as detailed in Supplementary Section A. To mitigate losses caused by the formation of Nb$_2$O$_5$, the qubits’ surfaces were capped with a $\sim$10\,nm thick layer of Gold~\cite{Grassellino2023}. This capping technique allowed us to achieve long $\text{T}_1$ values, ranging from tens up to hundreds of microseconds, an essential feature for observing relaxation time variations on microsecond timescales. 

\begin{table}[b]
    \centering
       \caption{Expected rate of interactions in the Sapphire substrate at the two experimental locations. The expected interaction rates were obtained by scaling the simulation results using both measured and theoretical inputs. Specifically, the $\gamma$-ray flux in the experimental rooms were measured with a 3” portable NaI spectrometer, yielding (1.7$\pm$0.9)\,$\gamma$/cm$^2$/sec at FNAL and (1.0$\pm$0.5)\,$\gamma$/cm$^2$/sec at LNGS. 
       For muons, a flux of 1\,$\mu$/cm$^2$/min was assumed at FNAL, based on the site altitude, while a six-order-of-magnitude suppression factor was applied to estimate the flux at LNGS~\cite{Aglietta1998}. Lastly, the radioactive contamination of the experimental setup components was measured in Ref.~\cite{cardani:2023}.}
\label{tab:RadioactivityBudget}
\begin{tabular}{lcc}
\hline
 Source             & FNAL                        & LNGS \\
                    & [events/sec]                & [events/sec] \\
\hline
Lab $\gamma$-ray   & $(31 \pm 2)\times10^{-3}$    & $(1.3 \pm 0.1)\times10^{-3}$ \\
Muons               & $(8 \pm 0.5)\times10^{-3}$   & $ < 10^{-5}$ \\
Setup               & $(2.7 \pm 0.5)\times10^{-3}$ & $(2.7 \pm 0.5)\times10^{-3}$\\
\hline
Total               &$(42 \pm 3)\times10^{-3}$      & $(4.0 \pm 0.6)\times10^{-3}$\\
\hline
\end{tabular}
\end{table}

To predict the rate of hits due to ionizing radiation, we performed a Monte Carlo simulation using the GEANT4 framework~\cite{AGOSTINELLI2003250} developed in Ref.~\cite{cardani:2023}. We anticipate the total rate of radiation events interacting within the chip at FNAL and LNGS to be $(42 \pm 3) \times 10^{-3}$\,events/sec and $(4.0 \pm 0.6) \times 10^{-3}$\,events/sec, respectively. Table~\ref{tab:RadioactivityBudget} summarizes the simulation results, with detailed information provided in Supplementary Section B. At FNAL (Fig.~\ref{fig:compare}(a), left panel), the rate of impacts in the Sapphire substrate is dominated by $\gamma$-rays from naturally occurring radioactive isotopes, with cosmic-ray muons also contributing significantly. At LNGS (Fig.~\ref{fig:compare}(a), right panel), muon interactions are reduced by six orders of magnitude due to the 1.4 km rock overburden, and gamma radiation is minimized by Copper and Lead shields, installed both inside and around the cryostat.
The last source of radiation arises from the materials of the chip and its surrounding components. In a prior study within the SQMS \emph{Round Robin} project~\cite{cardani:2023}, we measured the radioactive content of each component and determined that cables, connectors, amplifiers, and circulators contribute negligibly to the overall rate. However, the PCBs, located near the chip, exhibits significant radioactive contamination. This contribution dominates the ``Setup” rates in Table~\ref{tab:RadioactivityBudget} and, while negligible at FNAL, becomes the primary source at LNGS where external radiation is highly suppressed.

To explore the impact of elevated radiation levels, at LNGS, we used calibrated Thorium sources to increase the event rate up to one event every two seconds. This allowed us to investigate transmons' behavior in a controlled “high” radiation environment.

\section{Qubit relaxation time in different radiation environments}
\label{sec:comparison}
When a particle impinges on the qubit chip, it releases energy into the substrate. Muons create long tracks across the chip, while $\gamma$-rays interact through photo-electric absorption or (mainly via) Compton scattering, producing short-track electrons. In the absence of an electric field, the thousands of charges created along the ionizing track recombine into phonons, which diffuse throughout the chip~\cite{Wilen:2021}. As illustrated in Fig.~\ref{fig:protocol}(a), these phonons can break Cooper pairs in superconductors, creating quasiparticles.
When these quasiparticles tunnel through the Josephson Junction, they may cause the qubit to lose energy and  decay to its ground state $\ket{g}$~\cite{martinis:2021}, inducing prolonged relaxation periods and significantly reducing $\text{T}_1$. Previous studies indicated that these low $\text{T}_1$ periods can last from one to several tens of milliseconds~\cite{Li2024direct, McEwen:2021}. 

Transmons with energy relaxation rates of 1/40 and 1/32\,$\mu$s$^{-1}$ were already characterized in a facility in which radioactivity could be controlled by using a movable lead shield. The authors of the paper, determined that environmental radioactivity was negligible for qubits with that $\text{T}_1$~\cite{Vepsalainen:2020}. In this work, we characterized the 8 qubits and chose to focus on qubit ``Q1" (Supplementary Section A), which exhibited the best combination of T$_1 \ (\sim80 \ \mu$s) and readout fidelity.

We performed standard $\text{T}_1$ measurements in both above-ground (FNAL) and underground (LNGS) environments using the same qubit.
Each data point required about 50 seconds to record, as can be seen in Fig.~\ref{fig:compare}(b). The results showed similar mean values and fluctuations in $\text{T}_1$, consistent with typical transmon behavior~\cite{Grassellino2023}, and no abrupt $\text{T}_1$ drops were observed. This outcome aligns with expectations, as standard $\text{T}_1$ experiments, due to their averaging nature, lack the temporal resolution necessary to capture millisecond-scale fluctuations. 
As a consequence, to probe such rapid events, we developed a tailored fast decay detection protocol~\cite{McEwen:2021, mcewen2024gap_engineering} specifically designed to monitor single qubits with sub-millisecond resolution (Section~\ref{sec:detection protocol}).
Nevertheless, the standard $\text{T}_1$ experiment was performed multiple times: initially at FNAL, then at LNGS, and finally again at FNAL, interleaved with the measurements described in the next section. This allowed us to verify that the qubit behavior remained stable and reproducible throughout the entire data-taking campaign. 

\section{Detection protocol}
\label{sec:detection protocol}
\begin{figure}[t]
\includegraphics[width=\columnwidth]{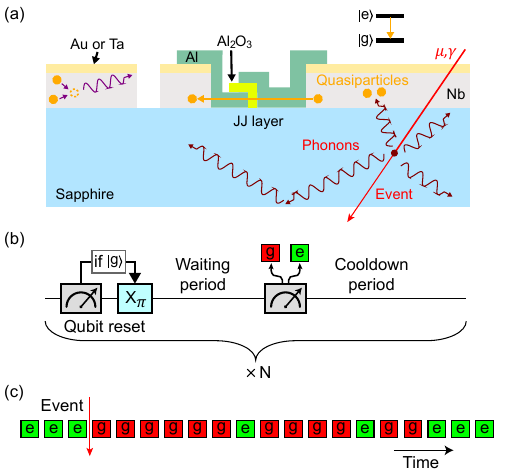}
\caption{\textbf{Quasiparticle generation process and experimental protocol.} (a) Schematic representation of quasiparticles generation 
due to particle interactions in the substrate. Ionizing radiation impinging on the substrate produces electron-hole pairs that recombine, creating phonons. These phonons spread throughout the chip and, in superconducting materials, can break Cooper pairs into quasiparticles, which can tunnel across the Josephson Junction inducing decay of the qubit from the excited state $\ket{e}$ to the ground state $\ket{g}$. (b) The fast decay detection protocol consists of repeated cycles of qubit preparation in the excited state, a waiting period, measurement, and cooldown period. Each cycle lasts up to 74\,$\mu$s, with waiting times of 5\,$\mu$s. (c) During normal operation, the qubit is likely to remain in the excited state due to its long $\text{T}_1$. Radiation events drastically reduce $\text{T}_1$, causing repeated detections in $\ket{g}.$ Over time, the qubit gradually recovers its natural $\text{T}_1$. These sequences enable the detection of radiation impacts with sub-millisecond resolution.}
\label{fig:protocol} 
\end{figure}

The detection protocol (Fig.~\ref{fig:protocol}(b)) begins by resetting the qubit to its first excited state $\ket{e}$ using a conditional $\pi$-pulse, applied after an initial measurement if the qubit is found in the ground state $\ket{g}$. After a waiting period $\Delta t_d$, the qubit state is measured again, followed by a cooldown period to avoid populations in higher energy states. Specifically, the waiting period was fixed at 5 \,$\mu$s, with readout pulse durations varying from 5 to 9\,$\mu$s. The $\pi$-pulse duration was negligible ($\sim200$\,ns), and cooldown periods were set between 24 and 58\,$\mu$s. These parameters were chosen dataset by dataset to optimize both state initialization fidelity and readout efficiency. 
The total time to complete a cycle can be thought as a sampling period (T$_S$). Measurements of Q1 were done with  T$_S$ increasing from 40 to 74\,$\mu$s. 

Under normal conditions, the qubit predominantly remains in $\ket{e}$ due to its relatively long $\text{T}_1$, with occasional decays to $\ket{g}$ occurring at a probability of $P(g) = (1-e^{-\Delta t_d/\text{T}_1})\approx \Delta t_d/\text{T}_1$. Radiation events, however, significantly reduce $\text{T}_1$, leading to a much higher probability of detecting the qubit in $\ket{g}$. Since the effect of quasiparticles lasts longer than the duration of a single cycle, an event due to radiation typically manifests as consecutive detections in $\ket{g}$, as illustrated in Fig.~\ref{fig:protocol}(c).

Data were collected using an RFSoC board equipped with the Quantum Instrumentation Control Kit (QICK)~\cite{stefanazzi2022qick}. Detailed setup information, including filtering and amplification stages, are provided in Supplementary Section F. Data corresponding to qubit Q1 were collected both underground at LNGS in 2023 (where the chip was also exposed to radioactive sources) and at FNAL in 2025.
At FNAL, we additionally investigated the impact of different sampling periods on the reconstruction of radiation events. To assess the stability of the results over time, we also acquired datasets on different days using identical experimental parameters. A summary of the acquired runs is reported in Table~\ref{tab:ds}.

\section{Identifying Candidate Radiation Events from Qubit Measurements}
\label{sec:candidate events}
Each run typically lasted several hours, during which hundreds of millions of readouts were recorded.
Each readout generates I/Q (in-phase and quadrature-phase) data, which are post-processed to identify the qubit state.  
By plotting a large number of measurements on the I-Q plane,  two main clusters corresponding to the $\ket{g}$ and $\ket{e}$ states were identified.
Occasionally, signals associated with the qubit being in the second excited state $\ket{f}$ or, less frequently, in the third excited state $\ket{h}$ were also observed.

The data were segmented into traces of $10^6$ measurements and, for each trace, the observed clusters were fitted with 2D Gaussian functions to estimate the population in each state. Traces exhibiting elevated populations in the $\ket{f}$ or $\ket{h}$ states were excluded from the analysis to maintain data consistency. While transitions involving $\ket{f}$ or $\ket{h}$ could offer valuable insights for future studies, the current methodology prioritizes a robust and reproducible dataset to facilitate comparison across different runs. This approach ensures a high-purity signal sample, but reduces the effective live-time of the measurements (Table~\ref{tab:ds}).

\begin{figure}[t]
\includegraphics[width=\columnwidth]{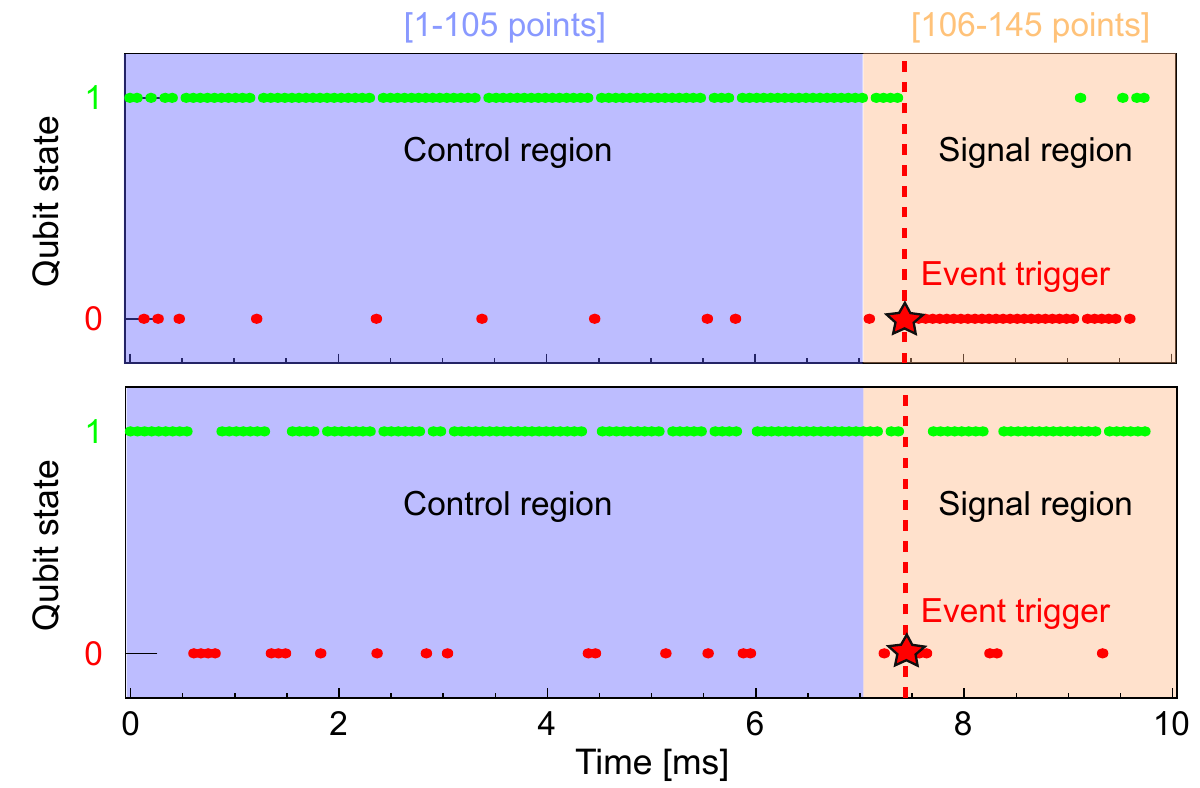}
\caption{\textbf{Examples of triggered events.} The top panel shows a clear radiation-induced event, identifiable by a significant excess of zeros in the signal region relative to the control region. In contrast, the bottom panel displays a triggered event discarded due to noise, as indicated by the high number of zeros already present in the control region. These examples demonstrate the importance of using the control region to filter out noise fluctuations and isolate true radiation-induced events.}
\label{fig:TriggeredEvents} 
\end{figure}

After this filtering process, the clusters corresponding to $\ket{g}$ and $\ket{e}$ were rotated to align their centers along the same Q value, enabling straightforward state discrimination by applying a threshold on the I axis (further details on the methodology are provided in Supplementary Section C). This threshold was used to convert the measured I/Q values into a binary sequence of 0s (for $\ket{g}$) and 1s (for $\ket{e}$).

Since the two clusters were not fully separated, any chosen threshold introduced some errors in state identification. To minimize these errors, the clusters were fitted with 1D Gaussian functions to calculate the fraction of misidentified measurements for each state. By varying the threshold value, we found the one that minimizes the total misidentification rate (namely the sum of the two fractions). This approach achieved a correct state identification rate of 90–95\%, depending on the specific dataset.

For each trace of 10$^6$ events (corresponding to approximately one minute of data), we obtained a sequence of 0s and 1s. To disentangle the 0s produced by radioactivity from those caused by spontaneous qubit decay, we required the detection of at least four consecutive zeros to trigger an event. This criterion reduces false triggers due to spontaneous decay and other noise sources such as electronic noise, vibrations~\cite{Kono_2024,Yelton2025} and material relaxation at low temperatures~\cite{Adari_2022,Chang2025}. These effects are particularly relevant at the interface between the holder   and the chip~\cite{anthonypetersen2022, Angloher_2023, Armengaud_2016}, or in the metallic films~\cite{romani2024, Angloher_2023}. We also repeated the analysis using a trigger threshold of three consecutive zeros and obtained consistent results.

Once an event was triggered, a 145-point window was recorded (see Fig.~\ref{fig:TriggeredEvents}) and divided into two regions:

\begin{itemize}
\item \textbf{Control region} ($[1, 105]$ points): used to estimate the baseline number of zeros (noise) before the event;
\item \textbf{Signal region} ($[106, 145]$ points): captured the event, including 5 points before and 35 after the trigger.
\end{itemize}

The length of the Signal region was chosen through an iterative process. We initially assumed a signal duration of 100 points and performed the analysis described later in the text. Since no events with more than 60 zeros were observed, we reduced the signal window to 60 points. This adjustment preserves the number of zeros generated by radioactivity while reducing the number of zeros from spontaneous qubit decay within the same window. Again, we observed no events with more than 40 zeros in a 60-point window. We therefore further reduced the signal region to 40 points, which was found to be the optimal length to fully capture radiation-induced events while suppressing background noise from qubit decay.
To avoid re-triggering, a dead time of 35 points was applied after each trigger. Figure~\ref{fig:TriggeredEvents} illustrates two examples of triggered events, one of which was discarded due to not meeting relevant criteria detailed in the next section.

\section{Event Selection}
\label{sec:selection}
According to the Monte Carlo simulation summarized in Table~\ref{tab:RadioactivityBudget}, the event rates due to radioactivity are 0.004 and 0.042\,events/s at LNGS and FNAL, respectively.
These values assume that all events interacting with the substrate are successfully detected by the qubit. In practice, however, some detection inefficiency is expected, leading to lower actual rates.

During all Q1 measurements, we observed a ground-state population $P(g)$ ranging from  11.8 to 17.6$\%$ (Table~\ref{tab:ds}). A rough estimate suggests that, with this value and a trigger condition requiring four consecutive ground-state measurements, the expected trigger rate due to spontaneous qubit decay is approximately proportional to $P(g)^4$/T$_S$ resulting in few events per second (T$_S$ being the sampling period, or the time to complete a cycle).
Such a high noise level would completely obscure the 10$^{-3}$\,events/s generated by radioactivity.
Therefore, to isolate radiation-induced events after triggering, we apply additional data selection.

Our approach is based on computing the probability of observing more than a certain number of zeros in the signal region. Given a $P(g)$ of about 10$\%$, the most likely values of zeros in a 40-points window would be 3--5 (Binomial distribution). Higher numbers of zeros become less and less probable. Since we expect a rate from radioactivity of few 10$^{-3}$\,events/s, we decided to require the minimum number of zeros in the signal region (N$_{\rm signal}^{\rm min}$)  that ensures a noise from qubit decay  $< 1\times$10$^{-4}$\,events/s, thus an order of magnitude below the searched signal. The precise value of N$_{\rm signal}^{\rm min}$ in the signal region depends on the value of $P(g)$ and the sampling period $T_S$ for that specific run. Since this parameter is rather stable, the values of N$_{\rm signal}^{\rm min}$ to accept an event resulted 21-22 in a 40-points region for most of the Q1 datasets (Table~\ref{tab:ds}).

Similarly, the control region was examined to identify transient noise.
Events with an anomalous number of zeros in the control region -indicative of transient noise- were flagged and discarded, as they can lead to false triggers.
Specifically, we computed the binomial probability of observing a given number of zeros in the control region and rejected events with a probability lower than 1$\%$.
Since both an excess and a deficit of zeros can result in such low probabilities, we selected only events with a number of zeros between N$_{\rm control}^{\rm min}$ and N$_{\rm control}^{\rm max}$ reported in Table~\ref{tab:ds}.
The cut was intentionally chosen to be very loose, as its purpose was solely to remove events in which qubit performance deviated significantly from the average.
Detailed selection criteria for each dataset are provided in Table~\ref{tab:ds}.

The top panel of Fig.~\ref{fig:TriggeredEvents} shows a radiation-induced event, where the signal region exhibits a clear excess of zeros compared to the control region. In contrast, the bottom panel displays an event that was subsequently discarded. These examples highlight the role of the data selection in distinguishing genuine radiation-induced events from noise.

\begin{figure}[t]
\includegraphics[width=\columnwidth]{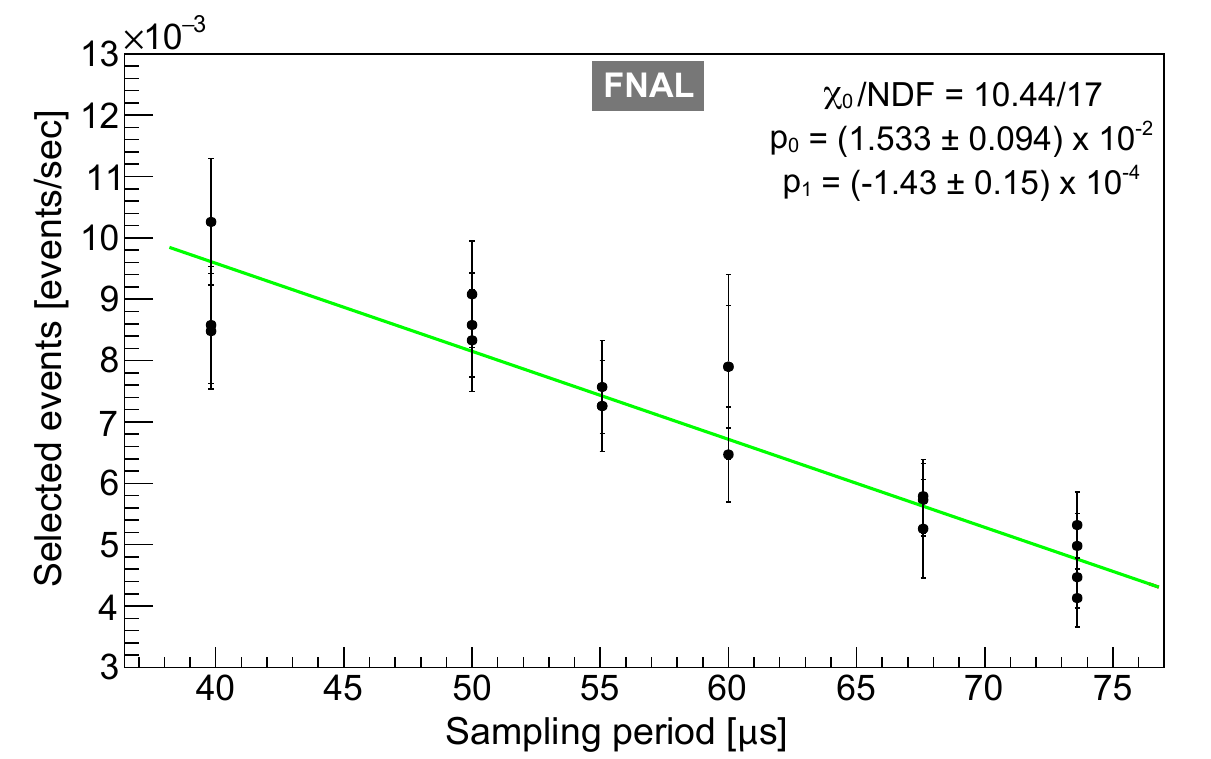}
\caption{\textbf{Measured rate of events for FNAL-Q1 as a function of the sampling period.} Data were modeled using a first-degree polynomial. The faster the sampling rate, the higher the detection efficiency.}
\label{fig:samplingperiod} 
\end{figure}

\section{Results and Discussion}
\label{sec:results}

The rates of events that passed the data selection procedure (Section~\ref{sec:selection}), as well as the specific parameters of each run, are listed in Table~\ref{tab:ds}.
The runs are reported following the same order in which data were acquired. Each run lasted typically four hours and we often took runs in different days to monitor the stability of the results.

\begin{table*}[t]
    \centering
    \caption{\textbf{List of datasets.} Data collected at LNGS and at FNAL. The table shows both the original acquisition duration and the ``live-time" duration after excluding intervals with high $\ket{f}$ population. The improvement in the detection protocol increased the live-time of the FNAL measurements of Q1. Multiple datasets were acquired using the same parameters to assess the reproducibility of the results. Columns T$_S$ and $P(g)$ report, respectively, the sampling period and the probability of detecting the qubit in the $\ket{g}$ state, computed from the ratio of measured 0s over the total records. N$_{\rm signal}^{\rm min}$ is the minimum number of zeros in the Signal region. N$_{\rm control}^{\rm min}$ and N$_{\rm control}^{\rm max}$ are the minimum and maximum number of zeros in the Control region. The last column shows the rate of events that passed the data selection.}
    \label{tab:ds}
      \begin{tabular}{@{}lccccccc@{}}
        \toprule
        \textbf{Datasets} & \textbf{Acquisition Time} & \textbf{Live-time } & \textbf{T$_S$}     & \textbf{$P(g)$} & \textbf{N$_{\rm signal}^{\rm min}$}  & \textbf{N$_{\rm control}^{\rm min, max}$}   & \textbf{Measured Rate}  \\
                          & \textbf{[min]}            & \textbf{[min]}      & \textbf{[$\mu$s ]} & \textbf{[$\%$]} &               &             & \textbf{[events/sec]} \\
        \midrule
        \multicolumn{7}{l}{\textbf{LNGS}} \\
        \midrule
        Shielded chip             & 736.7 & 209.96    &73.6 &11.8 &19 &6, 21 &(0.40 $\pm$ 0.18)$\times$10$^{-3}$\\
 \hline
        44 kBq Th Source          & 620.7 & 114.0     &67.6 &14.5 &21 &8, 24 &(16.0 $\pm$ 1.5)$\times$10$^{-3}$\\
        76 kBq Th Source          & 124.1 & 46.3      &67.6 &13.8 &21 &7, 23 &(16.6 $\pm$ 2.5)$\times$10$^{-3}$\\
        125 kBq Th Source         & 62.1  & 23.7      &67.6 &15.5 &22 &9, 25 &(26.8 $\pm$ 4.4)$\times$10$^{-3}$\\
        161 kBq Th Source         & 62.1  & 62.1      &67.6 &14.9 &21 &8, 25 &(29.4 $\pm$ 2.8)$\times$10$^{-3}$\\
      \midrule
        \multicolumn{7}{l}{\textbf{FNAL}} \\
        \midrule
        60\,$\mu$s - run1         &240.0  &55.0     &60.0 &14.7 &21 &8, 23 &(7.9 $\pm$ 1.5)$\times$10$^{-3}$\\
        60\,$\mu$s - run2         &240.0  &180.0    &60.0 &15.8 &22 &9, 24 &(6.47 $\pm$ 0.77)$\times$10$^{-3}$\\
        60\,$\mu$s - run3         &130.0  &90.0     &60.0 &15.5 &22 &9, 24 &(7.9 $\pm$ 1.0)$\times$10$^{-3}$\\
\hline
        68\,$\mu$s - run1         &270.4  &270.4    &67.6 &14.2 &21 &8, 23 &(5.26 $\pm$ 0.80)$\times$10$^{-3}$\\
        68\,$\mu$s - run2         &270.4  &270.4    &67.6 &14.3 &21 &8, 24 &(5.73 $\pm$ 0.59)$\times$10$^{-3}$\\
        68\,$\mu$s - run3         &270.4  &270.4    &67.6 &13.5 &20 &7, 23 &(5.79 $\pm$ 0.60)$\times$10$^{-3}$\\
 \hline
        74\,$\mu$s - run1         &294.4  &294.4    &73.6 &15.1 &21 &9, 25 &(5.32 $\pm$ 0.55)$\times$10$^{-3}$\\
        74\,$\mu$s - run2         &294.4  &294.4    &73.6 &14.5 &21 &8, 24 &(4.47 $\pm$ 0.50)$\times$10$^{-3}$\\
        74\,$\mu$s - run3         &294.4  &294.4    &73.6 &14.2 &21 &8, 23 &(4.98 $\pm$ 0.53)$\times$10$^{-3}$\\
        74\,$\mu$s - run4         &294.4  &294.4    &73.6 &15.0 &21 &8, 24 &(4.13 $\pm$ 0.48)$\times$10$^{-3}$\\
\hline
        50\,$\mu$s - run1         &200.0  &200.0    &50.0 &16.0 &22 &9, 26 &(8.33 $\pm$ 0.83)$\times$10$^{-3}$\\
        50\,$\mu$s - run2         &200.0  &200.0    &50.0 &16.3 &22 &9, 26 &(8.58 $\pm$ 0.85)$\times$10$^{-3}$\\
        50\,$\mu$s - run3         &200.0  &200.0    &50.0 &17.6 &23 &11, 28 &(9.08 $\pm$ 0.87)$\times$10$^{-3}$\\
\hline
        40\,$\mu$s - run1         &159.2  &159.2    &39.8 &15.8 &22 &9, 25 &(8.48 $\pm$ 0.94)$\times$10$^{-3}$\\
        40\,$\mu$s - run2         &159.2  &159.2    &39.8 &15.7 &22 &9, 25 &(8.58 $\pm$ 0.95)$\times$10$^{-3}$\\
        40\,$\mu$s - run3         &159.2  &159.2    &39.8 &15.0 &22 &8, 24 &(10.3 $\pm$ 1.0)$\times$10$^{-3}$\\
\hline
        55\,$\mu$s - run1         &220.4  &220.4    &55.1 &15.9 &22 &9, 26 &(7.26 $\pm$ 0.74)$\times$10$^{-3}$\\
        55\,$\mu$s - run2         &220.4  &220.4    &55.1 &16.2 &22 &9, 26 &(7.57 $\pm$ 0.76)$\times$10$^{-3}$\\
        55\,$\mu$s - run3         &220.4  &220.4    &55.1 &16.2 &22 &9, 26 &(7.26 $\pm$ 0.74)$\times$10$^{-3}$\\
 
         \bottomrule
    \end{tabular}
\end{table*}

First, we draw the reader’s attention to the ``FNAL Q1" data, which were acquired using various sampling periods ranging from 40 to 74\,$\mu$s. We observe that runs with the same sampling period yield consistent results, despite being recorded on different days or with slightly varying qubit lifetimes and hence different values of $P(g)$. Moreover, the data show that shorter sampling periods result in higher measured event rates. Specifically, using a sampling period of 74\,$\mu$s yields an event rate, averaged over four runs, of $(4.68 \pm 0.26)\times 10^{-3}$\,events/sec. When the sampling period is reduced to 40\,$\mu$s, the event rate, averaged again across the three runs, increases to $(9.08 \pm 0.56)\times 10^{-3}$\,events/sec. This trend is also illustrated in Fig.~\ref{fig:samplingperiod}, where the event rate is reported as a function of the sampling period.

\begin{figure}[t]
\includegraphics[width=\columnwidth]{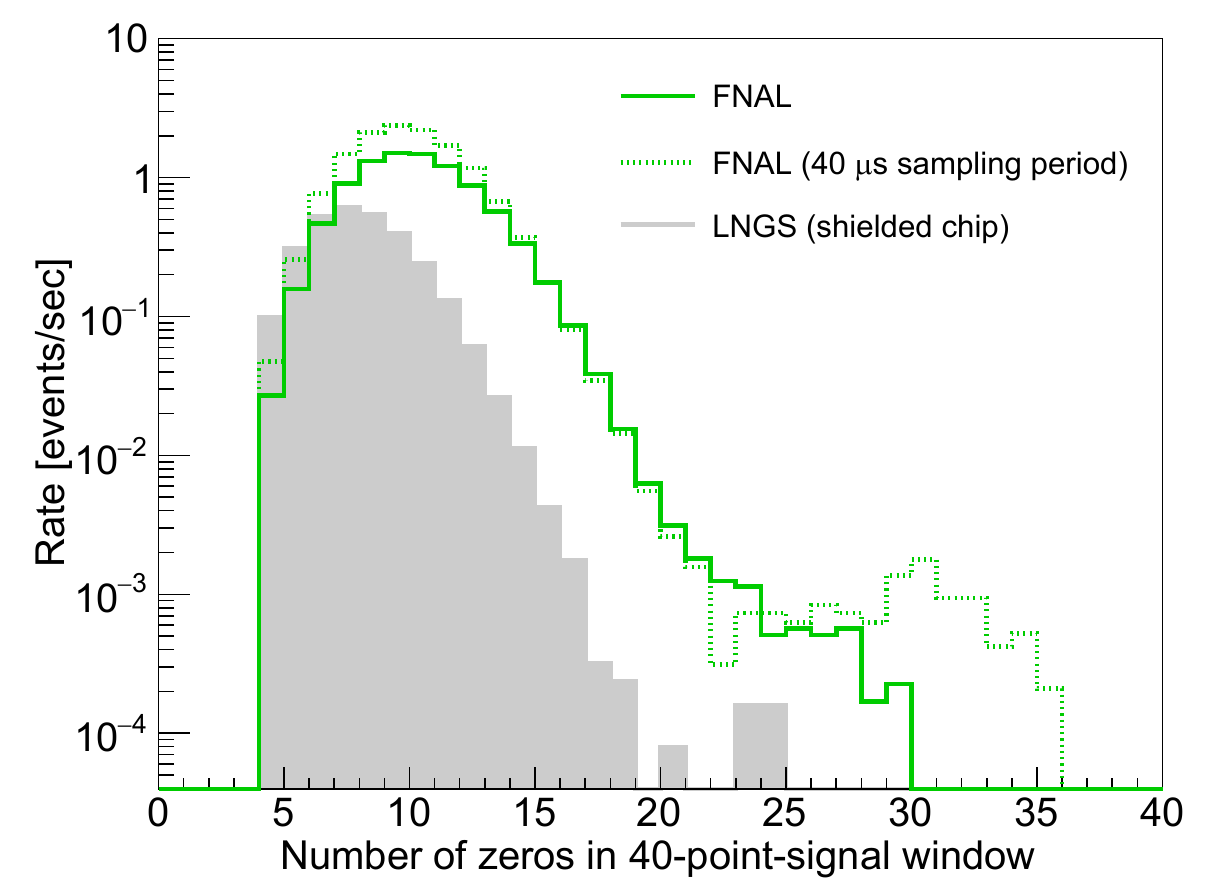}
\caption{\textbf{Distribution of zeros in the signal region for runs conducted deep-underground at LNGS (gray, filled histogram) and above-ground FNAL (green lines).} At FNAL, the qubits were completely unshielded from cosmic and ambient gamma radiation resulting in significantly more detected events. The dotted line represent FNAL data acquired with a sampling period of 40\,$\mu$s, the other FNAL run and the LNGS one were acquired with 74\,$\mu$s.}
\label{fig:ZerosSignalRegionBkg} 
\end{figure}

Such behavior was anticipated. Indeed, for each run, we chose selection criteria that ensured a qubit decay-induced noise rate below $10^{-4}$\,events/s, thus keeping the noise rejection efficiency constant. Since the qubit parameters were reasonably reproducible across all measurements, the chosen threshold was more or less the same---typically between 19 and 22 zeros. 
In contrast, the \emph{signal efficiency} increased with faster sampling periods. For example, if a signal lasts approximately 1.5 milliseconds, a sampling period of 40\,$\mu$s would produce about 37 zeros, while a 74\,$\mu$s sampling period would yield only 20. Given that the analysis threshold was set between 19 and 22 zeros (depending on the run), it is likely that events producing only 20 zeros on average have a low signal efficiency. Conversely, events producing 37 zeros would almost always pass the data selection. 
This effect can be appreciated in Fig.~\ref{fig:ZerosSignalRegionBkg}, where the two green histograms represent two runs acquired at FNAL with  40\,$\mu$s and  74\,$\mu$s sampling period. In the 40\,$\mu$s run there is a clear excess of events with a large number of 0s. From this study, we conclude that only runs with the same sampling period can be compared.

We now focus on the difference in the event rates measured for Q1 at FNAL and at the underground LNGS laboratory. As previously explained, a meaningful comparison requires the use of the same sampling period. For this reason, we report in Fig.~\ref{fig:ZerosSignalRegionBkg} the histograms obtained at both sites utilizing a 74\,$\mu$s sampling period. 

At LNGS, the ``background" run was acquired exclusively with this sampling, yielding a rate of $(0.40 \pm 0.18)\times 10^{-3}$\,events/sec. Using the same acquisition parameters at FNAL, we measured an average rate of $(4.68 \pm 0.26)\times 10^{-3}$\,events/sec -approximately an order of magnitude higher.

We also note that both the obtained rates are an order of magnitude smaller compared to prediction of the Monte Carlo simulation (Table~\ref{tab:RadioactivityBudget}). This hints that the ``detection efficiency" for events produced by ionizing radiation of the protocol developed in this work is around $\sim$10$\%$.
The study presented in Fig.~\ref{fig:samplingperiod} indicates that the signal efficiency is limited to approximately 10$\%$ primarily due to the sampling period. By repeating the measurement with a shorter sampling period of 40\,$\mu$s (instead of 74\,$\mu$s), we nearly doubled the measured signal rate and, consequently, increased the efficiency to 19$\%$.
However, we empirically determined that significantly faster sampling rates tend to excite the qubit to its second or third excited states, resulting in non-robust and unreliable measurements (see Section~\ref{sec:comparison}).\\

\begin{figure}[t]
\includegraphics[width=\columnwidth]{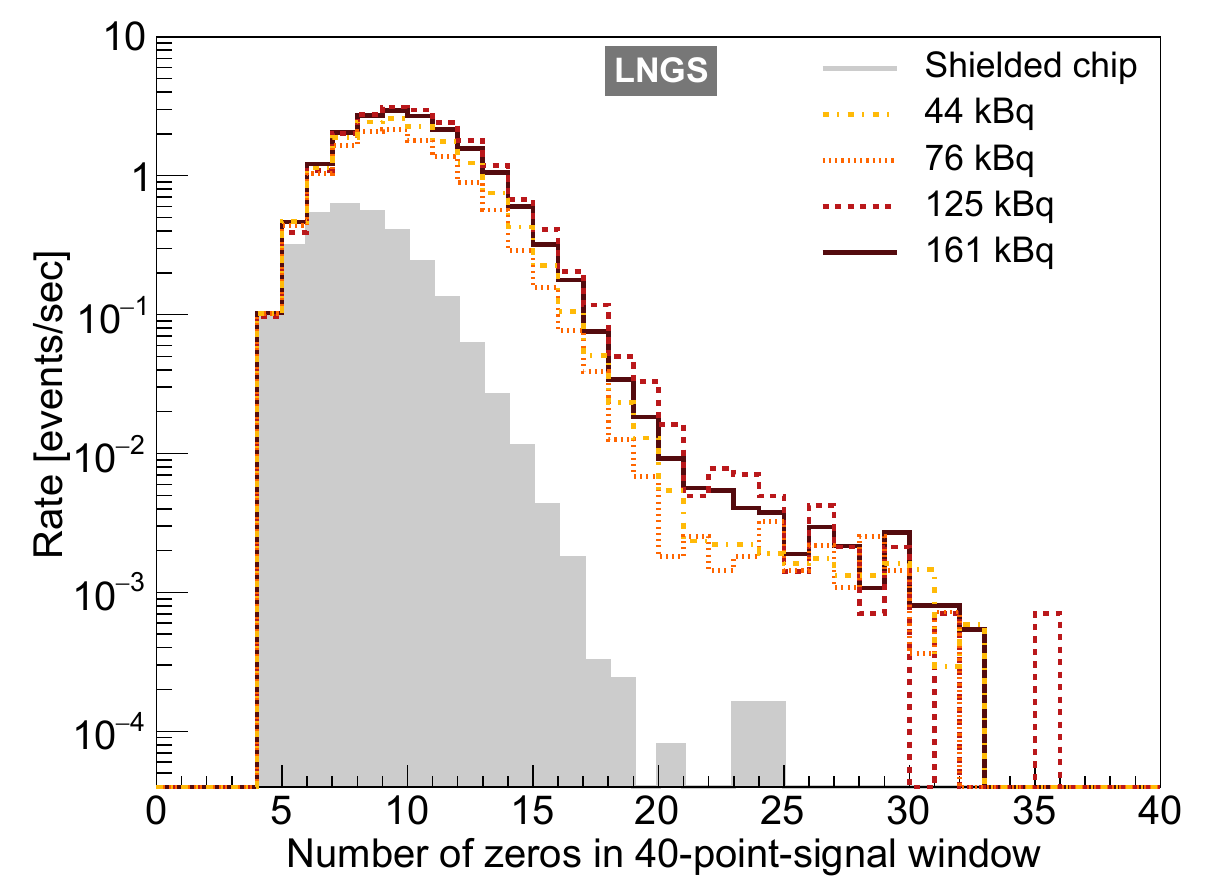}
\caption{\textbf{Distribution of zeros in the signal region for runs with controlled radioactive sources at deep-underground LNGS.} Data were acquired by exposing the chip to Thorium radioactive sources with increasing activity. For comparison, the distribution for the `background' run without sources is also shown (light gray). }
\label{fig:ZerosSignalRegionSources} 
\end{figure}

To further validate these conclusions, we investigated the detection efficiency for events due to ionizing radiation, at LNGS by exposing the chip to Thorium radioactive sources with increasing activity levels. For this study, commercially available Thorium rods with known activity were placed between the fridge's outer vacuum chamber and the external Copper shield. Data were analyzed using the same protocol applied to the background runs, and are presented in Fig.~\ref{fig:ZerosSignalRegionSources}. 

\begin{figure}[t]
\includegraphics[width=\columnwidth]{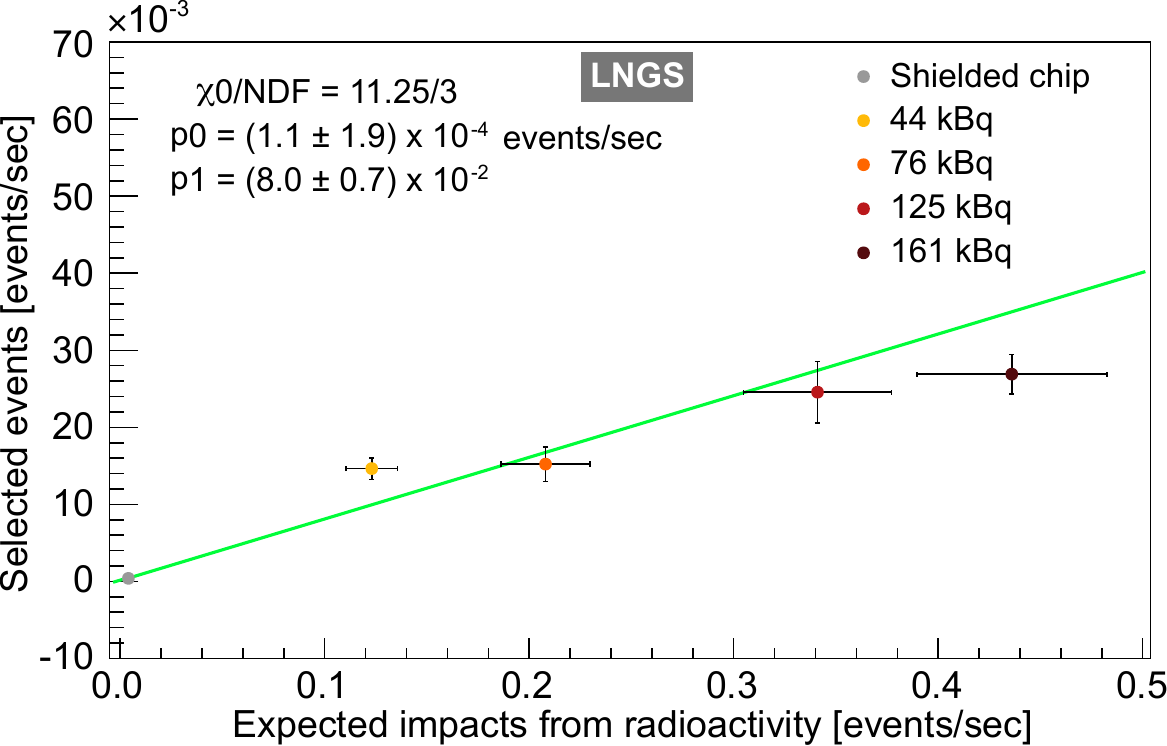}
\caption{\textbf{Measured rate of events for LNGS-Q1 as a function of the rate expected from the simulation.} Data were modeled using a first-degree polynomial. The linear coefficient p$_1$ is the ratio between the measured and expected events and can thus be considered as the detection efficiency. Statistical uncertainties are shown for the triggered rates, while uncertainties for expected impacts are dominated by systematic uncertainties in the simulation.}
\label{fig:scatter} 
\end{figure}

The measured rates indicate that, as expected, exposing the chip to a controlled radioactive source led to an excess of events with a high number of zeros in the signal region compared to the ``shielded" configuration. In Fig.~\ref{fig:scatter}, the rate of these events is reported as a function of the source’s intensity.
Since the runs with radioactive sources were done with a slightly faster sampling period, we corrected the rates reported in Table~\ref{tab:ds} by the effect of the sampling period (about 10$\%$). 
We acknowledge that comparing the background run and the runs with radioactive sources on the same plot is not fully justified. The presence of the sources could induce unknown effects on the qubit (indeed, the background run is the only one in which we observe such a low $P(g)$). Furthermore, the $\gamma$-ray sources exhibit a slightly different energy distribution compared to the background run, as shown in the plots provided in the supplementary material (\ref{sec:simulation}).
Nevertheless, this study shows that the rate of events attributed to radioactivity increases roughly linearly with the strength of the source. 

Additionally, we compared the triggered and simulated rates for the various Thorium sources. The slope parameter p$_1$ of the linear fit in Fig.~\ref{fig:scatter} indicates that the aggressive data selection results into a detection efficiency of $(8.0 \pm 0.7)\%$ (with the uncertainty being purely statistical), in qualitative agreement with our previous result.

Despite the low detection efficiency, our findings demonstrate that qubits are sensitive to gamma radiation, even though $\gamma$-rays deposit less energy compared to cosmic rays~\cite{Wilen:2021,Cardani:2021}. This result is the first experimental confirmation of the studies conducted in~\cite{Fink:2024}.
Apart from the consideration of that paper, we highlight that if qubits were to be used as particle detectors, $P(g)$ should be suppressed in order to relax the analysis threshold and, at the same time, the sampling frequency should be made faster.

Finally, we recall that even triggering on 4 consecutive zeros we obtained a rate of few events/second of single qubits errors, thus orders of magnitude higher compared to  the noise induced by radioactivity. This conclusion aligns with recent studies~\cite{Harrington:2024}, which report that cosmic rays are not the predominant source of the most frequent correlated errors among qubits. The observed radiation-like-induced events could therefore be linked to other mechanisms yet to be fully understood. Investigating these alternative potential sources of correlated errors will be a crucial focus of future studies, aiming to further elucidate the interplay between environmental noise and qubit performance.

\section{Conclusions}
We conducted a comparative study to assess the impact of radiation on transmon qubits in two laboratories with very different radiation environments. Standard $\text{T}_1$ measurements on the same qubit revealed no discernible differences in energy relaxation times, which remained consistently around 80 microseconds.

To probe time dynamics at shorter timescales, we developed a fast decay detection protocol with active reset tailored for a single qubit. The same qubit measured with this technique at FNAL (above ground and without Lead and Copper shielding) showed about 10 fold excess rate compared to the maximally shielded qubit at the deep-underground Gran Sasso Laboratory, as predicted by radioactivity studies. However, disentangling such a small excess rate from the single qubit noise required very stringent trigger and analysis thresholds, confirming that radioactivity is not the major source of errors in modern transmon qubits with relaxation times of the order of 0.1 milliseconds. 
Experiments with controlled Thorium sources at LNGS demonstrated the potential for detecting $\gamma$-ray impacts, opening the possibility of using transmons as particle detectors. However, the observed rates are lower than those predicted by simulations, indicating the need for more refined trigger algorithms and/or analysis strategies.

Future work will aim to achieve a deeper understanding of the error sources that currently overshadow radiation-induced effects. This will involve refining fast decay detection and analysis protocols, testing diverse qubit geometries and materials, and conducting simultaneous measurements across both the same and multiple chips. 
By addressing these challenges, we will move closer to realizing the full potential of transmon qubits in practical quantum technologies.

\acknowledgments
The authors thank the Director and technical staff of the Laboratori Nazionali del Gran Sasso. We are also grateful to the LNGS Computing and Network Service for computing resources and support on U-LITE cluster at LNGS. We thank  S.~Nisi for the ICP-MS measurements of the PCB, C.~Tomei and G.~D'Imperio for the maintenance of the simulation software, G.~Catelani for the useful discussion.
We are grateful to A.~Girardi for the controller of the cryogenic switch, M.~Guetti for the help with the whole cryogenic facility (including its continuous upgrades), and M.~Iannone for the assembly of the device at LNGS. We are thankful to Theodore C. III and Grzegorz Tatkowski for providing support on experimental preparations and cryogenic operations at FNAL.

The work was supported by the U.S. Department of Energy, Office of Science, National Quantum Information Science Research Centers, Superconducting Quantum Materials and Systems (SQMS) Center under the contract No. DE-AC02-07CH11359, by the Italian Ministry of Foreign Affairs and International Cooperation, grant number US23GR09, by the PNRR MUR project CN00000013-ICSC, by the Italian Ministry of Research under the PRIN Contract No. 2020h5l338 (“Thin films and radioactivity mitigation to enhance superconducting quantum processors and low temperature particle detectors”) and
by the Italian Ministry of Research under the PRIN Contract No. 2022BP4H73 (Chasing Light Dark Matter with Quantum Technologies”).

\bibliography{main}

\vspace{12pt}

\noindent
\textbf{Author Contributions}

\noindent
M.B. and F.C. designed and fabricated the transmon devices. S.Z. performed initial characterization of the qubits. A.C. T.R., F.D.D. and D.H. run the experiments.  C.B., N.C., F.D.D.,  A.M., A.R. and L.C. performed the data analysis. D.V.Z. helped in writing acquisition software. A.M. performed GEANT4 simulations. I.C., A.P., L.P. and S.P. oversaw the cryostat upgrade for the measurements and the cryogenic operations. T.R., A.M., F.D.D., and L.C. wrote the manuscript with inputs from all the co-authors in interpreting the results.  T.R., A.G. and L.C. conceived and supervised the whole project.

\newpage
\clearpage

\section*{SUPPLEMENT}
\beginsupplement

\subsection{Device Fabrication}
\label{supp:fab}

\begin{figure}[h]
\includegraphics[width=.8\columnwidth]{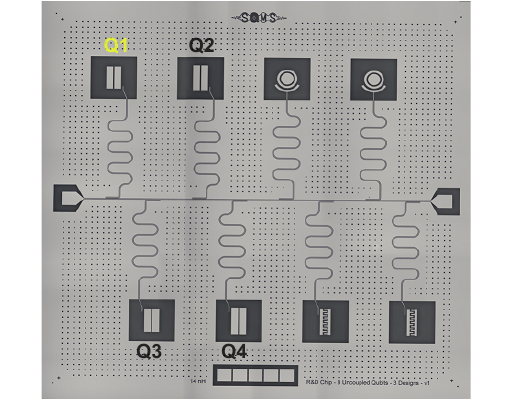}
\caption{\textbf{Chip layout.} Q1 was chosen for the measurements as it exhibited the best combination of $\text{T}_1$ ($\sim$80\,$\mu$s) and readout fidelity.}
\label{fig:Chip} 
\end{figure}

The device was fabricated on a 432 $\mu$m thick $c$-plane HEMEX grade double-polished Sapphire substrate with 2-inch diameter. Substrate preparation, base-layer (Nb) deposition, and Josephson junction evaporation followed the fabrication methods described in Ref.~\cite{Grassellino2023}. To prevent oxidation of the Niobium, a $~10$\,nm-thick Gold capping layer was deposited using the same sputtering tool without breaking the vacuum. 

Figure~\ref{fig:Chip} illustrates the chip layout. 
Qubits with concentric and inter-digitated pads are not used in the experiments due to their lower $\text{T}_1$ values. Q1, ... , Q4 were characterized and Q1 having a transition frequency of 4717.4\,MHz was chosen for the measurements as it exhibited the best combination of $\text{T}_1$ ($\sim$80\,$\mu$s) and readout fidelity.
The readout frequency was 7206.8\,MHz and the readout length varied between 5 to 9~$\mu$s.

\subsection{Simulation of Interaction Rates}
\label{sec:simulation}

\begin{figure}
\includegraphics[width=\textwidth]{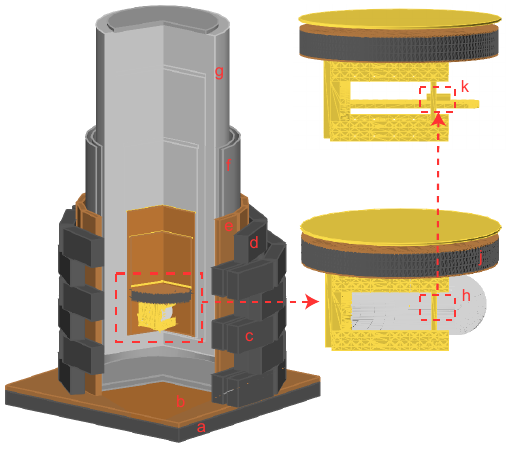}
\caption{\textbf{Experimental setup at LNGS as implemented in the simulation.} (a-e) External gamma shield; (f) external two-layer Permimphy magnetic shield; (g) cryostat with internal vessels; (h) internal CryoPerm® magnetic shield; (j) internal gamma shield; (k) gold-plated copper box hosting the chip. Component details are provided in the text.}
\label{fig:Sim_Geometry} 
\end{figure}

We used an updated version of the GEANT4-based Monte Carlo simulation described in Cardani et al.~\cite{cardani:2023} to predict both the rate of impacts and the energy deposited in the Sapphire substrate. The implemented geometry, shown in Fig.~\ref{fig:Sim_Geometry}, includes the following components (from outermost to innermost):

\begin{enumerate}
    \item [a.] An 80 $\times$ 80 cm$^2$, 5 cm-thick Lead base;
    \item [b.] An 80 $\times$ 80 cm$^2$, 5 cm-thick Copper base;
    \item [c.] Half a round of 10 $\times$ 20 cm$^2$, 5 cm-thick Lead bricks;
    \item [d.] One full round of 10 $\times$ 20 cm$^2$, 5 cm-thick Lead bricks;
    \item [e.] One full round of 10 $\times$ 66.5 cm$^2$, 2 cm-thick Copper bars;
    \item [f.] A two-layer Permimphy magnetic shielding (Outer layer: 55.6 cm (OD) $\times$ 91.3 cm (height), 1.5 mm-thick; Inner layer: 51 cm (ID) $\times$ 89 cm (height), 1.5 mm-thick);
    \item [g.] A simplified cryostat model with internal vessels (48.1 cm (OD) $\times$ 119.5 cm (height));
    \item [h.] A CryoPerm® magnetic shield located below the mixing chamber;
    \item [j.]  A 3 cm-thick Lead plate at 10 mK between the sample and electronic components, serving as an internal gamma shield;
    \item [k.] A gold-plated copper box (30.23 $\times$ 30.23 $\times$ 6.3 mm$^3$) housing the chip.
\end{enumerate}

The chip substrate (7.5 $\times$ 7.5 mm$^2$, 432 $\mu$m-thick HEMEX Sapphire) is the sole active volume, where we store tracks and record the energy released by each simulated interaction. The main external contributions, namely environmental gammas and muons, are generated according to the methods described in our previous work~\cite{cardani:2023}. For both FNAL and LNGS, interaction rates in the chip are obtained by scaling the number of recorded events to flux measurements detailed in Section 3 of Ref.~\cite{cardani:2023}, with the exception of the gamma flux at FNAL that was measured to be (1.7 $\pm$ 0.9)\,$\gamma$/cm$^2$/sec.

For close sources of radioactivity, we simulated radioactive decays of relevant isotopes ($^{40}$K, $^{232}$Th and $^{238}$U chains) uniformly distributed within the volume of the two PCBs (RO4003C produced by ROGERS Corporation) located at the opposite ends of the qubit chip. PCBs, indeed, constitute the only relevant contribution from close sources, as outlined in Ref.~\cite{cardani:2023}. We then estimated the interaction rates in the chip by scaling the number of recorded events to the activities reported in Table~\ref{tab:RadioactivityMeasPCB}. These activities were determined by converting the concentrations measured using Inductively Coupled Plasma Mass Spectrometry (ICP-MS) at the LNGS Chemistry facility.

\begin{table}[b]
    \centering
       \caption{\textbf{Activities of relevant radioactive isotopes in RO4003C PCBs produced by ROGERS Corporation.} These values were determined by converting the corresponding concentration measured by ICP-MS at the LNGS. Chemistry facility.}
\begin{tabular}{lccc}
\hline
 Source &$^{40}$K   &$^{232}$Th & $^{238}$U\\
        &$\big[$mBq/kg$\big]$   & $\big[$mBq/kg$\big]$  & $\big[$mBq/kg$\big]$\\
 \hline
    PCB &(518.8 $\pm$ 259.4)    &(164.0 $\pm$ 41.0)  &(161.2 $\pm$ 49.6)\\
\end{tabular}
\label{tab:RadioactivityMeasPCB}
\end{table}

\begin{figure}[t]
\includegraphics[width=\columnwidth]{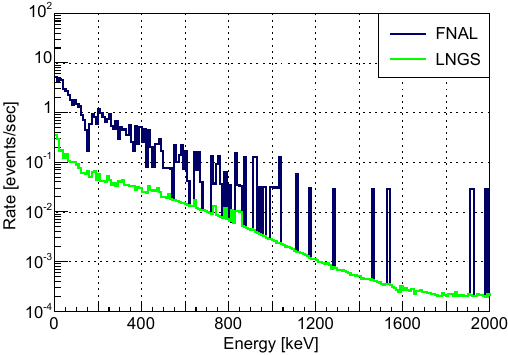}
\caption{\textbf{Rate of interactions expected in the Sapphire chip as a function of the deposited energy.} The results are shown for both FNAL (blue) and LNGS (green). FNAL rate is dominated by environmental gammas ($\sim 42 \times 10^{-3}$ events/sec), while at LNGS the primary contribution comes from radioactive decays in the PCBs ($\sim 2.7 \times 10^{-3}$ events/sec).}
\label{fig:RateSimulation} 
\end{figure}

Fig.~\ref{fig:RateSimulation} presents the results of the Monte Carlo simulation for both FNAL and LNGS facilities, showing the rate of events as a function of the energy deposited in the qubit substrate. The event rates for each radiation source are reported in Table~\ref{tab:RadioactivityBudget}.
Notably, the overall rate is expected to be approximately an order of magnitude lower at LNGS (4 $\times 10^{-3}$ event/sec) than at FNAL (42 $\times 10^{-3}$ event/sec), primarily due to lower contributions from external sources.

\begin{figure}[t]
\includegraphics[width=\columnwidth]{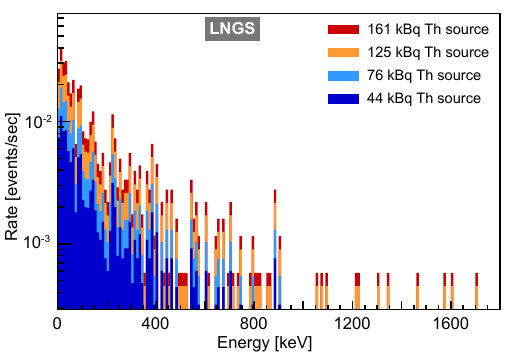}
\caption{\textbf{Interaction rates expected in the Sapphire substrate as a function of the energy deposited in the chip when exposed to Thorium radioactive sources.} The distinct colors in the plot correspond to the various activities used in the measurements.}
\label{fig:RateThSourceSimulation} 
\end{figure}

To estimate interaction rates in the presence of Thorium radioactive sources at LNGS, we generated radioactive decays from a point-like $^{232}$Th source positioned between the external magnetic shield and the Copper bars, approximately at the same height of the chip. Only gamma emitters in the chain ($^{228}$Ra, $^{228}$Ac, $^{212}$Pb, $^{212}$Bi and $^{208}$Tl) are included, since $\alpha$ and $\beta$ particles are shielded by the cryostat. The interaction rates in the chip are determined by scaling the number of recorded events to the activities reported in Table~\ref{tab:RateThSource_Simulation}, which correspond to the Thorium source activities employed in the measurements (evaluated using High-Purity Germanium spectroscopy). Results are shown in Fig.~\ref{fig:RateThSourceSimulation}.

\begin{table}[b]
    \centering
       \caption{\textbf{Interaction rates expected in the sapphire substrate for different Thorium source activities.} The values are obtained by scaling the Monte Carlo simulation to match the activities used in the measurements. Errors include both the statistical uncertainty and a 5\% systematic uncertainty associated with the Monte Carlo simulation.}
\begin{tabular}{lc}
\hline
Activity &Rate \\
$\big[$kBq$\big]$ &$\big[$events/sec$\big]$ \\
\hline
    44.2 &0.12 $\pm$ 0.01 \\
    75.9 &0.20 $\pm$ 0.02 \\
    125.4 &0.34 $\pm$ 0.04 \\
    161.0 &0.43 $\pm$ 0.05 \\    
\end{tabular}
\label{tab:RateThSource_Simulation}
\end{table}

\subsection{Qubit State Discrimination}
\begin{figure*}[t]
    \centering
    \includegraphics[width =\textwidth]{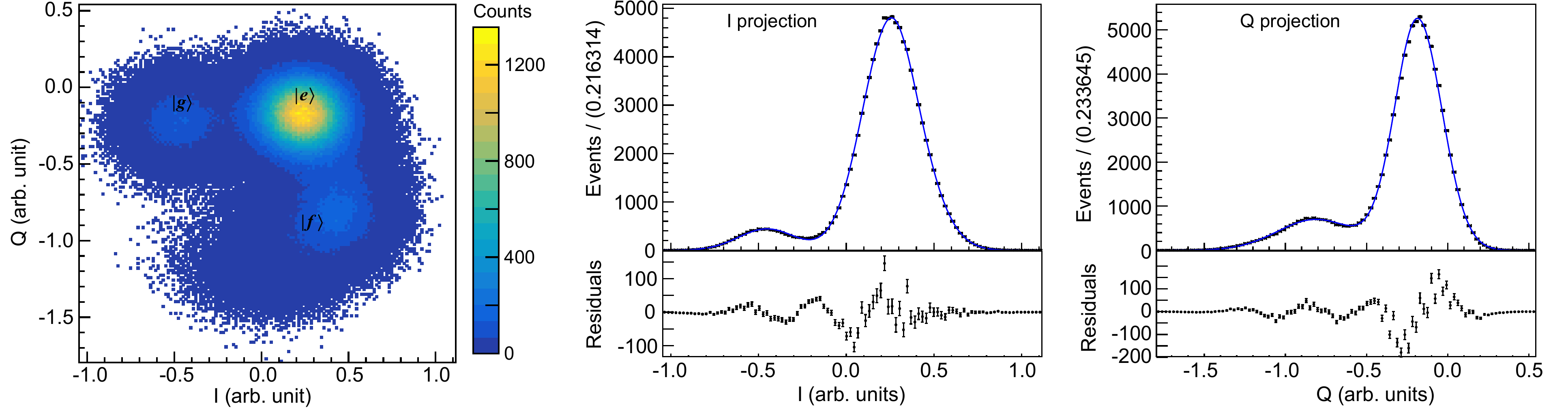}
    \caption{\textbf{Clouds on the I-Q plane corresponding to the qubit states for a subset of data.} (Left) The clouds corresponding to $\ket{g}$ (left), $\ket{e}$ (upper right), and $\ket{f}$ (bottom right) are visible. (Center and Right) Fits performed on the projected clouds to estimate state populations.}
    \label{fig:Clouds}
\end{figure*}

The I/Q values (in-phase and quadrature-phase signals) obtained from each measurement reflect the qubit state. When plotted on the I-Q plane, these values form clusters (or clouds) corresponding to $\ket{g}$ (ground), $\ket{e}$ (1st excited), and occasionally higher states like $\ket{f}$ (2nd excited) and $\ket{h}$ (3rd excited). 
To ensure robustness in the procedure, measurements with excessive populations of $\ket{f}$ or $\ket{h}$ are excluded from the analysis.

To identify and exclude such periods, we fit the clusters with 2D Gaussian functions, one for each state. Our data are divided into traces of $10^6$ measurements, with fits performed separately on each trace using the RooFit toolkit~\cite{RooFit}. The clouds are projected along the I and Q axes, and both projections are simultaneously fitted. The fit function for each projection is $\sum_{i}A_i \cdot f\big(x, \bar{x}_i, \sigma_i \big)$, where $f$ is a Gaussian centered at $\bar x_i$ with standard deviation $\sigma_i$, and $A_i$ is a normalization factor that represents the number of measurements for state $i$. The state population is then given by $P_i = A_i / \sum_i A_i$.

Fig.~\ref{fig:Clouds} refers to a single trace, and shows the clouds and the fitted projections used to estimate state populations. The $\ket{g}$ and $\ket{e}$ clusters are aligned along the Q axis, allowing discrimination between the two states using a threshold only on the I axis, simplifying the identification process.

\subsection{``Effective" $\text{T}_1$ Calculation}
To calculate the ``effective" relaxation time $\text{T}'_1$ over a trace of $10^6$ points, we perform the following steps:

\begin{enumerate}
    \item Count the occurrences of the qubit in the excited state $\ket{e}$ ($\text{N}_e$) relative to the total number of measurements, ($\text{N}_{\rm tot} = 10^6$);
   \item Compute $\text{T}’_1$ using the formula: 
    \begin{equation}
        \text{T}'_1 = -\frac{\Delta t_d}{\text{log}(\text{N}_e/\text{N}_{\rm tot})}.
    \end{equation}
\end{enumerate}

This process is repeated for each segment of $10^6$ points to track variations over time. Finally, we calculate the average $\langle\text{T}’_1\rangle$, which serves to estimate the probability of the qubit being in state $\ket{g}$ for each dataset (Table~\ref{tab:ds}).

\begin{figure}[b]
    \centering
    \includegraphics[width=\columnwidth]{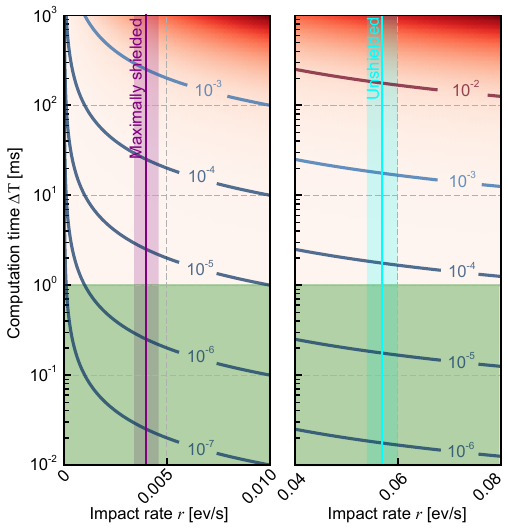}
    \caption{\textbf{Relationship between available computational time and radiation rate for different error probability tolerance.} The contours show fixed error probability lines as a function of available computational time $\Delta$T and the rate of impacts $r$. The purple (cyan) vertical line with shaded area represents  the simulated rate for maximally shielded (LNGS) versus non-shielded condition. The horizontal green area ($\Delta$T $<$ 1 ms) highlights the lifetimes of modern transmon qubits. Error rates in this regime are below $10^{-4}$, making radiation an insignificant contributor.}
    \label{fig:P_Impact}
\end{figure}
 
\subsection{Impact of Radiation on Computational Timescales}
\begin{figure*}[t]
\includegraphics[width=\textwidth]{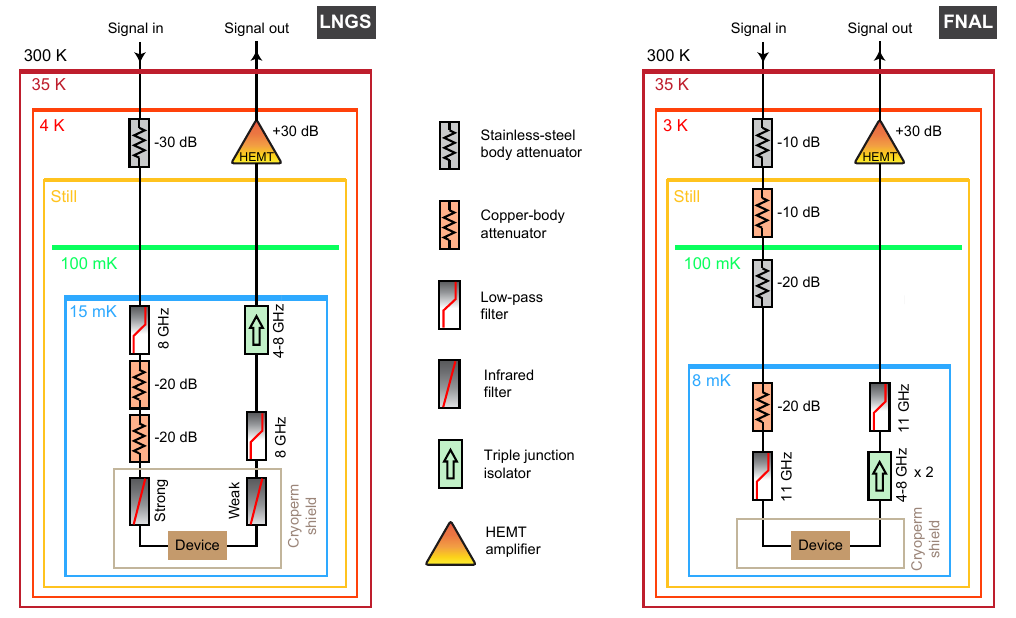}
\caption{\textbf{Cryogenic wiring diagram of the measurement setups at LNGS and at FNAL.}}
\label{fig:wiring} 
\end{figure*}

We utilized the simulated rates to provide an upper estimate for the timescales at which radiation would start to affect quantum computation through information loss occurring on individual qubits. The probability of at least one event occurring in a time window $\Delta$T is given by $P_{\rm impact} = 1 - e^{-r\cdot\Delta\text{T}}$, where $r$ is the rate of impacts.
In a non-shielded laboratory setting, radioactivity affects quantum computations less than 0.1\% (1\%) of the time if the computation is completed within approximately $17\,\mathrm{ms}$ ($170\,\mathrm{ms}$). Note that the computation may include multiple runs of an algorithm and not necessarily require to be a single run lasting for $\Delta$T. In a fully shielded environment like LNGS, due to a much lower expected radiation rate, we have instead $P_{\rm impact} < 0.1\%$ (1\%) if $\Delta \text{T} \lesssim 250\,\mathrm{ms}$ (2.5 s). The available computation time increases further if a higher error probability is tolerated. Fig.~\ref{fig:P_Impact} shows the available computational time as a function of radiation impact rates around values estimated for the two cases. The contour lines represent various constant error rates, and the horizontal green shaded area ($<1$ ms) corresponds to the $\text{T}_1$ times of contemporary transmon qubits. In this regime, radiation induces errors with a negligible $P_{\rm impact} < 10^{-4}$. 
This means that even if a calculation is repeated over $10^4$ times, at most one result would be corrupted due to radiation.

Most modern quantum algorithms, such as the variational quantum eigensolver or quantum approximate optimization, require less than a millisecond for a single iteration and involve a large number of repetitions ($\sim\mathcal{O}(10^4)$). Therefore, we can conclude that environmental radiation has a negligible effect on current transmon-based quantum computing platforms, even if error correction fails due to a radiation impact. However, as ongoing research continues to extend coherence times~\cite{Grassellino2023}, radiation-induced errors will become a significant concern for large-scale quantum processors. Conventional error-correcting codes, such as surface codes~\cite{Fowler2012surface_code}, which rely on detecting and correcting uncorrelated errors in nearby qubits, may struggle to address these errors, as radiation events often disrupt multiple adjacent qubits simultaneously~\cite{McEwen:2021}. To overcome this challenge, novel error-correcting codes like quantum low-density parity-check (LDPC) codes~\cite{Breuckmann2021LDPC} are being developed. These codes leverage long-range interactions, making them more resilient to localized errors caused by radiation and other environmental factors~\cite{Kono_2024}.

\subsection{Measurement Setup}
\label{sec:setup}
The measurement setups at the two locations are shown in Fig.~\ref{fig:wiring}. The transmon qubit is measured in a hangar geometry, where a $\lambda/4$ coplanar waveguide (CPW) readout resonator is coupled to a transmission line. Both qubit control and readout pulses are sent through a common input line. This input line is attenuated by 60-70 dB, with attenuation distributed between the 4-K and mixing chamber stages. At the base plate, copper-body 20 dB attenuators are used to ensure effective thermalization. 

At LNGS, the input line is additionally filtered through a 8 GHz reflective low-pass filter and a QMC-CRYOIRF-002 infrared filter to suppress high-frequency noise. The superconducting output line is protected from the incoming noise through a 4-8 GHz triple junctions isolator, an 8 GHz low-pass filter, and a weak infrared filter (QMC-CRYOIRF-001). At FNAL, both the input and output lines are equipped with 11 GHz reflective low-pass filters. The superconducting output line also has two 4-8 GHz isolators. 

To shield the device boxes from stray magnetic fields, CryoPerm® enclosures are used at both locations.  The output signal is first amplified by a commercial high-electron-mobility transistor (HEMT) amplifier at the 4-K stage, followed by a room-temperature amplifier outside the fridge before digitization. The qubit control and readout pulses are digitally generated using an RFSoC ZCU216 evaluation board and the QICK software suite~\cite{stefanazzi2022qick}.

\end{document}